\def\year{2016}\relax
\documentclass[letterpaper]{article}
\usepackage{aaai16}

\usepackage{times}

\usepackage{graphicx}
\usepackage{amssymb,amsmath,bm}
\usepackage{textcomp}
\usepackage{booktabs}
\usepackage{multirow}
\usepackage[usenames,dvipsnames]{xcolor}
\usepackage{comment}
\usepackage{xcolor}

\usepackage{hyperref}

\usepackage{balance}  
\usepackage{graphics} 
\usepackage{times}    
\usepackage{url}      

\usepackage{times}
\usepackage{helvet}
\usepackage{courier}

\begin{document}
%
\title{Real-time On-Demand Crowd-powered Entity Extraction\\(Supplementary Technical Report)}
\author{
Ting-Hao (Kenneth) Huang\\
Carnegie Mellon University\\
Pittsburgh, PA, USA\\
{\tt tinghaoh@cs.cmu.edu}
\And
Yun-Nung (Vivian) Chen\\
National Taiwan University\\
Taipei, Taiwan\\
{\tt yvchen@csie.ntu.edu.tw}
\And
Jeffrey P. Bigham\\
Carnegie Mellon University\\
Pittsburgh, PA, USA\\
{\tt jbigham@cs.cmu.edu}
}

\maketitle

\noindent \textbf{[Note] This is the supplementary technical report of our paper ``Real-time On-Demand Crowd-powered Entity Extraction,'' which was published at the 5th Edition of The Collective Intelligence Conference (CI 2017) as an oral presentation~\cite{chen2017real}.} The original paper was only 3-page long, so we decided to share extra technical details in this report.

\vspace{.5pc}

\begin{abstract}



Modern dialog systems rely on accurate entity extraction to understand user utterances.
However, entity extraction is brittle due to data scarcity, language variability, and out-of-vocabulary entities.
To bridge this gap, we propose a real-time crowdsourcing solution based on the ESP game for image labeling.
When multiple players agree, entities can be reliably extracted from an utterance.
This approach is advantageous because it does not require training data.
Further, it is robust to unexpected input and capable of recognizing new entities.
Our approach achieves better F1-scores than that of the automated baseline for complex queries with a reasonable response time.
The proposed method is also evaluated via Google Hangouts' text chat and demonstrates the feasibility of real-time crowd-powered entity extraction.

\end{abstract}




\section{Introduction}

To understand user utterances, modern dialog systems rely heavily on entity extraction, known as the core task of \textit{slot filling} in many dialog system frameworks such as Olympus~\cite{Olympus}.
The goal of slot filling is to identify from a running dialog different \textit{slots}, which correspond to different parameters of the user's query.
For instance, when a user queries for nearby restaurants, key slots for \textit{location} and \textit{preferred food} are required for a dialog system to retrieve the appropriate information.
Thus, the main challenge in the slot-filling task is to extract the target \textit{entity}.

Dialog systems face three key challenges in entity extraction.
Due to \textbf{data scarcity}, labeled training data, which many existing technologies require to identify entities such as Conditional Random Fields (CRF)~\cite{AtisSpain,FeatureDropout2014} and Recurrent Neural Networks~\cite{Rnn2015}, are often unavailable for the wide variety of dialog system tasks.
Furthermore, it is more difficult to acquire the complicated conversational data required by other alternative dialog technologies, such as statistical dialog management~\cite{POMDPDia} or state tracking~\cite{StateTracking}.
Second, existing entity extraction technologies are not robust enough to identify \textbf{out-of-vocabulary entities}.
Even when labeled training data for the targeted slot could be collected,
state-of-the-art supervised learning approaches are brittle in extracting unseen entities.
\cite{FeatureDropout2014} find that the CRF-based entity extractor performed significantly worse when dictionary features were not used.
Third, challenges are also posed by \textbf{language variability}.
Successful applications process diverse input languages where potential entities are unlimited.
Therefore, to robustly serve arbitrary input, dialog systems must collect new sources of entities and update accordingly.

Research on dialog systems has focused on utilizing the Internet resource to extract entities such as movie names~\cite{wang2014leveraging};
Unsupervised slot-filling approaches have also been developed in recent years~\cite{UnsuperSlotFilling,UnsuperSlotFilling2}.
However, these methods are still underdeveloped.

\begin{figure}[t]
    \centering
    \includegraphics[width=\columnwidth]{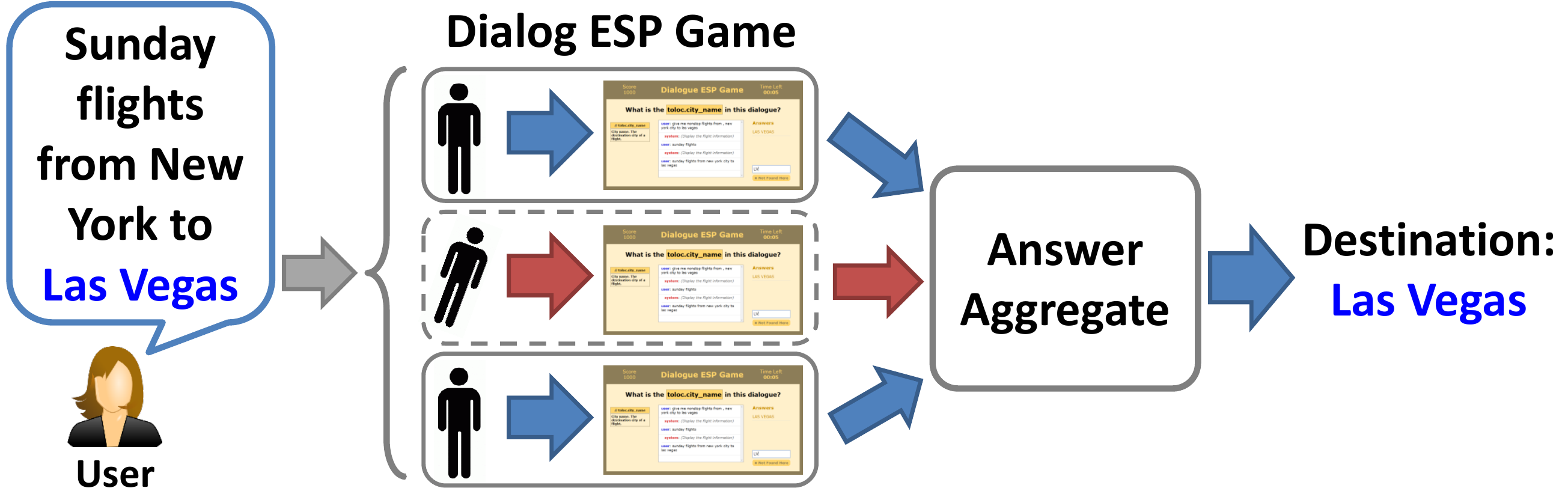}
    \caption{The crowd-powered entity extraction with a multi-player Dialog ESP Game.
By aggregating input answers from all players, our approach is able to provide good quality results in seconds. }
    \label{fig:flowchart}
\end{figure}

To address these challenges, we propose to use real-time crowdsourcing as an entity extractor in dialog systems.
To the best of our knowledge, few previous works have attempted to use crowdsourcing to extract entities from a running conversation.
\cite{wang2012crowdsourcing}, for example, studied various methods to acquire natural language sentences for a given semantic form by the crowd.
\cite{lasecki2013conversations} utilized crowdsourcing to collect dialog data, and illustrated CrowdParse, a system that uses the crowd to parse dialogs into semantic frames.
Recently, \cite{huang2015guardian} presented a crowd-powered dialog system called Guardian that uses the crowd to extract information from input utterances.
However, none of these works conducted formal studies on crowd-powered entity extraction in real-time.

Inspired by the ESP game for image labeling~\cite{EspGame2004}, we propose a \textbf{Dialog ESP Game} to encourage crowd workers to accurately and quickly perform entity extraction.
The ESP Game matches answers among different workers to ensure label quality, and we use a timer on the interface (Figure~\ref{fig:ui}) to ensure input speed.
Our method offers three main advantages: 1) it does not require training data; 2) it is robust to unexpected input; and 3) it is capable of recognizing new entities.
Furthermore, answers submitted from the crowd can be used as training data to bootstrap automatic entity extraction algorithms.
In this paper, we conduct experiments on a standard dialog dataset and user experiments with 10 users via Google Hangouts' text chatting interface.
Detailed experiments demonstrate that our crowd-powered approach is robust, effective, and fast.


In sum, the contributions of our work are as follows:

\begin{enumerate}
    \item We propose an ESP-game-based real-time crowdsourcing approach for entity extraction in dialog systems, which enables accurate entity extraction for a wide variety of tasks.
    \item To strive for real-time dialog systems, we present detailed experiments to understand the trade-offs between entity extraction accuracy and time delay.
    \item We demonstrate the feasibility of real-time crowd-powered entity extraction in instant messaging applications.
\end{enumerate}

\section{Real-time Dialog ESP Game}

\begin{figure}[t]
    \centering
    \includegraphics[width=0.99\linewidth]{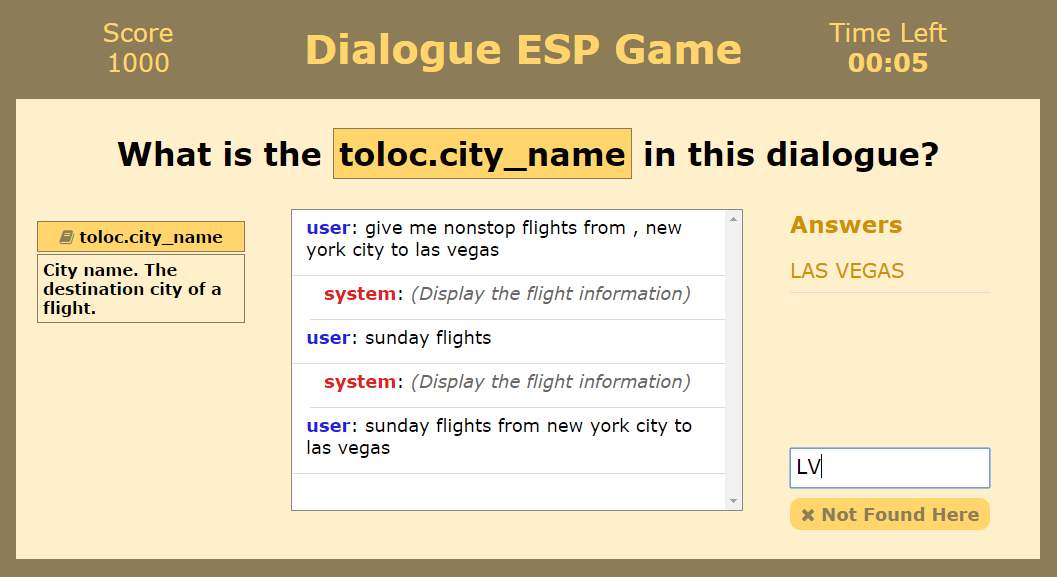}
    \caption{The Dialog ESP Game interface is designed to encourage quick and correct entity identification by crowd workers. Workers are shown the complete dialog and a description of the entity they should identify.}
    \label{fig:ui}
\end{figure}

We utilize real-time crowdsourcing with a multi-player Dialog ESP Game setting to extract the targeted entity from a dialog\footnote{The source code of worker interface and the data collected in Experiment 2 are available at:\\ \href{https://github.com/windx0303/dialogue-esp-game}{https://github.com/windx0303/dialogue-esp-game}}.
The ESP Game was originally proposed as a crowdsourcing mechanism to acquire quality image labels~\cite{EspGame2004}.
The original game randomly pairs two players and presents them with the same image.
Each player guesses the labels that the other player would answer.
If the players match labels, each is awarded 1000 points.
Our approach replaces the image in the ESP Game with a dialog chat log and players answer the required entity name within a short time.
We also relax the constraints of player numbers to increase game speed.
As Figure~\ref{fig:flowchart} shows, by aggregating input answers from all players, the Dialog ESP Game is able to provide high quality results in seconds.

Figure~\ref{fig:ui} shows the worker's interface.
When input dialog utterances reach the crowd-powered entity extraction component,
workers are recruited from crowdsourcing platforms such as Amazon Mechanical Turk (MTurk).
The timer begins counting down when the input utterance arrives, 
and the worker sees the remaining time on the top right corner of the interface (Figure~\ref{fig:ui}). When two workers match answers, a feedback notification is displayed, and the workers earn 1000 points. When the time is up, the task automatically closes.

To recruit crowd workers quickly, many approaches have been used in real-time crowd-powered systems such as VizWiz~\cite{VizWiz2010} and Chorus~\cite{Chorus2013}.
The \textit{quikTurkit} toolkit (quikturkit.googlecode.com) attracts workers by posting tasks and using old tasks to queue workers.
Similarly, the Retainer Model maintains a retaining pool of workers-in-waiting, who receive a signal when tasks become available.
Prior research shows that the Retainer Model is able to recall 75\% of workers within 3 seconds~\cite{CrowdIn2Sec2011}.
In Experiment 1, we first focus on the speed and performance of the Dialog ESP Game itself instead of recruiting time.
In Experiment 2, we propose a novel approach to recruit workers within 60 seconds and discuss details of the end-to-end response speed.

\section{Experiment 1:\\Applying Dialog ESP Game\\on ATIS Dataset}

To evaluate the Dialog ESP Game for entity extraction, 
we conducted experiments on MTurk to extract names of destination cities from a flight schedule query dialog dataset, the Airline Travel Information System (ATIS) dataset.

\paragraph{ATIS Dataset}
The ATIS dataset contains a set of flight schedule query sessions,
each of which consists of a sequence of spoken queries (utterances).
Each query contains automatic speech recognized transcripts and a set of corresponding SQL queries.
All queries in the data set are annotated with the query category: A, D, or X.
Class A queries are context-independent, answerable, and formed mostly in a single sentence;
however,
real-world queries are more complex.
In the ATIS data set, 32.2\% queries are context-dependent (Class D)
and 24.0\% of the queries are cannot be evaluated (Class X)~\cite{AtisCorpus1991}.
The ``context-dependent'' Class D queries require information from previous queries to form a complete SQL query.
For instance, in one ATIS session, the first query is ``From Montreal to Las Vegas'' (Class A).
The second query in the session is ``Saturday,'' which requires the destination and departure city name from the first query, and is thus annotated as Class D.
Class X is of all the problematic queries, e.g., hopelessly-vague or unanswerable. 

\paragraph{Data Pre-processing \& Experiment Setting}

For Class A, we obtain the preprocessed data used in many slot filling works~\cite{FeatureDropout2014,Rnn2015,AtisUk,AtisSpain,tur2010left},
which contain 4,978 queries for training,
893 queries for testing,
and 491 queries for developing.
200 queries are randomly extracted from the developing set for our study;
For Class D and X, 
we obtain the original training set of ATIS-3 data~\cite{Atis3Stats},
which contains 364 sessions and 3,235 queries.
200 Class-D queries are randomly selected from 200 distinct sessions.
For each extracted query, all previous queries before it within the same session
are also obtained
and displayed in the worker's interface (Figure~\ref{fig:ui}).
The same process is used to extract 150 Class-X queries for the experiments.
Note that in this work we focus only on the
\textbf{\texttt{toloc.city\_name}} slot (name of destination city),
which is the most frequent slot type in ATIS.
For each extracted query of Class D and X,
we define the last-mentioned destination city name of the flight in the query history (including the extracted query)
as the gold-standard slot value.


\begin{figure*}[t]
    \centering
    \includegraphics[width=0.99\linewidth]{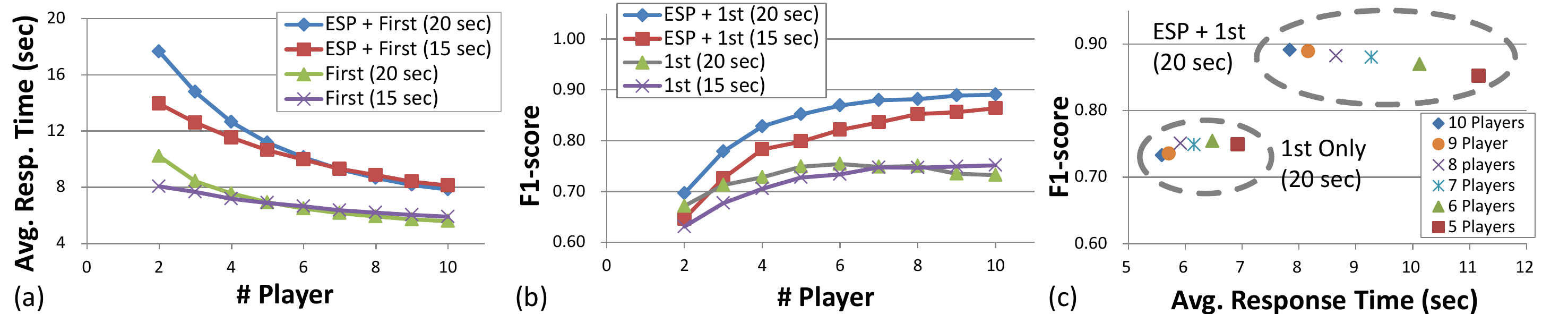}
    \caption{Trade-off curves between accuracy, average response time and number of players.}
    \label{fig:result}
\end{figure*}

\subsection{Understanding Accuracy and Speed Trade-offs}
\label{sec:study}

In order to design an effective crowd-powered real-time entity extraction system,
it is crucial to understand trade-offs between accuracy and speed.
These trade-offs correspond to the three main variables in our system:
the \textbf{number of players} recruited to answer each query in the Dialog ESP Game, 
the \textbf{time constraint} that each player has to answer a query,
and the \textbf{method to aggregate input answers}.
We have 3 ways to aggregate the input answers from the ESP game:

\begin{itemize}
    \item \textbf{ESP Only}:
        Return the first matched answer.
        If no answers match within the given time, return an empty label.
    \item \textbf{$i$th Only}:
        Return the $i$th input answer ($i$ = 1, 2, ...).
        For example, $i$ = 1 means to return the first input answer.
    \item \textbf{ESP + $i$th}:
        Return the first matched answers of the ESP game.
        If no answers match within the given time, return the $i$th answer.
\end{itemize}

We recruit 10 players for each ESP game, and randomly select player results to simulate the conditions of various player numbers.
All results reported in Experiment 1 are the averages of 20 rounds of this random-pick simulation process.
After empirically testing the interface, 
we run two sets of studies with time constraints set at 20 and 15 seconds, respectively.
Different methods to aggregate input answers could result in different response speed and output quality.
Note that if there are not any input answers, the methods above will wait until the time constraint and return an empty label.
In the actual experiments,
5 Dialog ESP Games for 5 different Class-A queries are aggregated in one task, with
an extra scripted game at the beginning as a tutorial.
When the first game ends,
the timer of the second ESP game starts and a browser alert informs the worker.
All experiments are run on MTurk;
800 Human Intelligence Tasks (HITs) are posted, and 588 unique workers participate in this study.

\begin{table}[t]
\small
\centering
\begin{tabular}{@{}ccccccc@{}}
\toprule
\begin{tabular}[c]{@{}c@{}}Time \\ Const. \end{tabular} & Aggregate  & \begin{tabular}[c]{@{}c@{}}\#\\ Player\end{tabular} & \begin{tabular}[c]{@{}c@{}}Avg. \\ Resp.\\ Time \end{tabular} & P & R & F1 \\ \midrule
\multirow{6}{*}{20s} & \multirow{2}{*}{\begin{tabular}[c]{@{}c@{}}ESP+\\ 1st\end{tabular}} & 10 & 7.837s & .867 & .916 & \textbf{.891} \\ \cmidrule(l){3-7} 
 &  & 5 & 11.160s & .828 & .877 & .852 \\ \cmidrule(l){2-7} 
 & \multirow{2}{*}{\begin{tabular}[c]{@{}c@{}}1st\\ Only\end{tabular}} & 10 & \textbf{5.590s} & .713 & .753 & .732 \\ \cmidrule(l){3-7} 
 &  & 5 & 6.924s & .730 & .769 & .749 \\ \cmidrule(l){2-7} 
 & \multirow{2}{*}{\begin{tabular}[c]{@{}c@{}}ESP\\ Only\end{tabular}} & 10 & 7.837s & .867 & .916 & \textbf{.891} \\ \cmidrule(l){3-7} 
 &  & 5 & 11.160s & .856 & .797 & .826 \\ \midrule
\multirow{6}{*}{15s} & \multirow{2}{*}{\begin{tabular}[c]{@{}c@{}}ESP+\\ 1st\end{tabular}} & 10 & 8.129s & .837 & .893 & .864 \\ \cmidrule(l){3-7} 
 &  & 5 & 10.628s & .799 & .798 & .798 \\ \cmidrule(l){2-7} 
 & \multirow{2}{*}{\begin{tabular}[c]{@{}c@{}}1st\\ Only\end{tabular}} & 10 & 5.895s & .739 & .764 & .751 \\ \cmidrule(l){3-7} 
 &  & 5 & 7.136s & .729 & .726 & .727 \\ \cmidrule(l){2-7} 
 & \multirow{2}{*}{\begin{tabular}[c]{@{}c@{}}ESP\\ Only\end{tabular}} & 10 & 8.129s & .860 & .865 & .863 \\ \cmidrule(l){3-7} 
 &  & 5 & 10.628s & .872 & .637 & .736 \\ \midrule
\end{tabular}
\caption{Dialog ESP Game results in Class A given different settings of
number of players, time constraint (Time Const.), and the method to aggregate input answers.}
\label{tab:classA}
\end{table}

Table~\ref{tab:classA} shows the results on Class A queries.
With 10 players and a 20-second time constraint,
the Dialog ESP Game achieves a best F1-score of 0.891
by the ``ESP+1st'' setting, and 
achieves the fastest average response time of 5.590 seconds 
by the ``1st'' setting.
The \textbf{ESP+1st} setting achieves the best F1-score,
and the \textbf{$1$st Only} setting has the shortest response time.
In most cases, tightening the time constraint provides a faster response but reduces output quality.

\begin{figure}[t]
    \centering
    \includegraphics[width=\columnwidth]{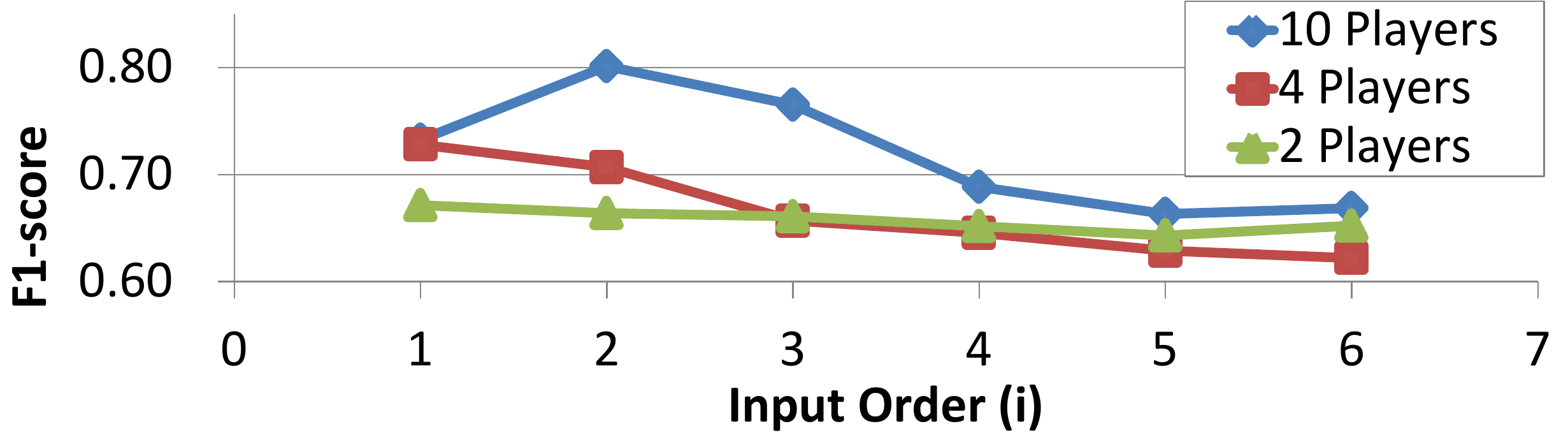}
    \caption{F1-score of the ``$i$th Only'' setting. Earlier input answers are generally of better quality (unless \#players $\geqslant$ 10, where almost all ESP games have at least one matched answer and the $i$th answer might not be solely used.)}
    \label{fig:order}
\end{figure}

We also analyze the relations among worker numbers, performance, and response time.
First, Figure~\ref{fig:order} shows output quality with respect to answer's input order.
On average, earlier input answers are of better quality, unless 10 or more players participate in the game.
However, with 10 players, almost all ESP games have at least one matched answer pair so that the $i$th answer is not solely used.
Therefore,
for the following experiments, we set $i$ as 1.
Second, in Figure~\ref{fig:result}(a) we observe the relations between the number of players and average response time.
Adding players reduces the average response time for all settings.
Third, the relations between number of players and output quality are also analyzed.
Figure~\ref{fig:result}(b) shows that the F1-scores increase when adding more players, even with the ``1st Only'' setting.
Finally, Figure~\ref{fig:result}(c) demonstrates the trade-offs between performance and speed.
For a fixed number of players, different input aggregate methods have different response times and F1-scores.
The ESP game requires more time for input answer matching, but in return output quality increases.

\subsection{Bootstrapping Automated System Performance}

\begin{figure}[t]
    \centering
    \includegraphics[width=\columnwidth]{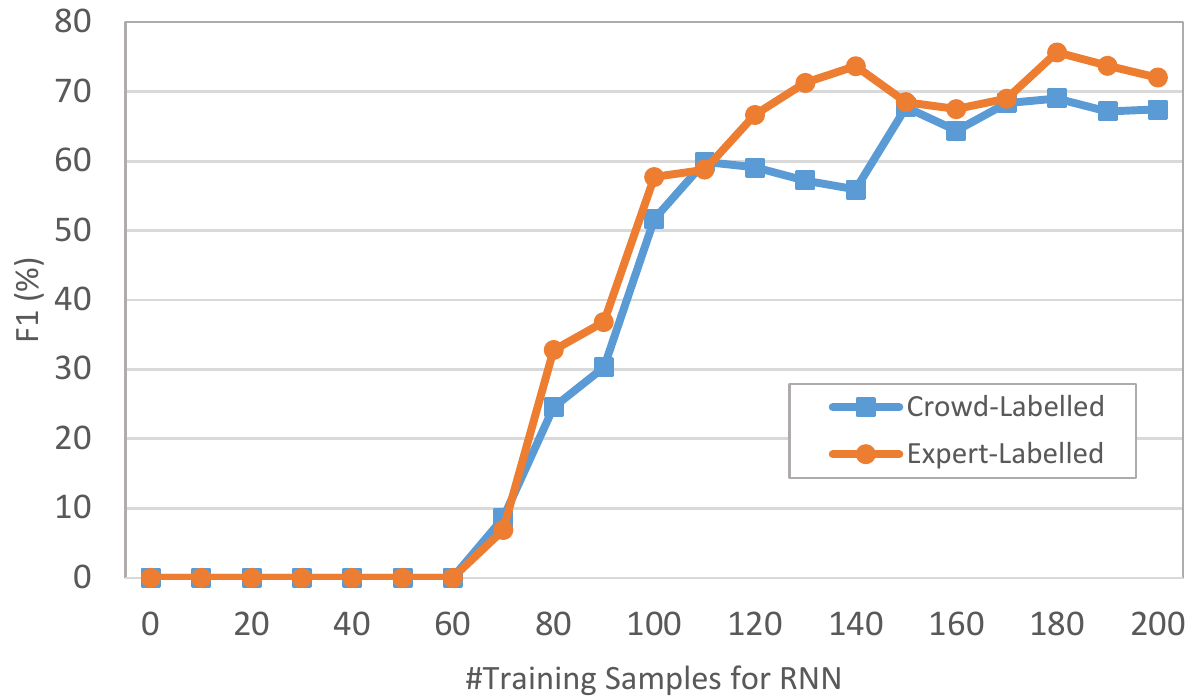}
    \caption{The performance of the extracted destination city predicted by the RNN-GRU trained on crowd-labelled and expert-labelled data. Crowd-labelled labels can be used to train machine-learning models and achieve comparable performance to expert-labelled data.}
    \vspace{-.5pc}
    \label{fig:rnn_curve}
\end{figure}

Once our crowd-powered system starts extracting entities, the collected annotations can serve as training data for the automated system. 
In order to see whether the crowd-annotated data is good enough for training a machine learning model, and how many of text instances are required to achieve reasonable performance, 
we train a state-of-the-art language understanding model, implemented by an recurrent neural network (RNN) with GRU cells~\cite{mesnil2015using,chen2016knowledge}, on the crowd-labelled or expert-labelled 200 ATIS conversations used in our study, and use the standard ATIS Class-A testing set to compare the performance in terms of the number of training samples.
Figure~\ref{fig:rnn_curve} illustrates the performance curves of models trained on crowd-labelled and expert-labelled data.
The result shows that crowd-generated labels can be used to train machine-learning models for bootstrapping in new domains, and achieve comparable performance to expert-labelled data.

\subsection{Evaluation on Complex Queries}

\begin{table*}[htpb]
\small
\centering
\begin{tabular}{|c||c|c|c|c||c|c|c|c||c|c|c|c|}
\hline
Query Category & \multicolumn{4}{c||}{\begin{tabular}[c]{@{}c@{}}Class D (context-dependent) \end{tabular}} & \multicolumn{4}{c||}{\begin{tabular}[c]{@{}c@{}}Class X (unevaluable) \end{tabular}} & \multicolumn{4}{c|}{\begin{tabular}[c]{@{}c@{}}Class A (context-independent) \end{tabular}} \\ \hline
Methods & \begin{tabular}[c]{@{}c@{}}Resp. Time\end{tabular} & P & R & F1 & \begin{tabular}[c]{@{}c@{}}Resp. Time\end{tabular} & P & R & F1 & \begin{tabular}[c]{@{}c@{}}Resp. Time\end{tabular} & P & R & F1 \\ \hline
CRF Baseline & 0.043s &  .776 &  .307 &  .440 &  0.061s &  .636 &  .285 &  .393 &  0.019s &  .985 &  .987 &  \textbf{.986} \\ \hline
1st Only & 5.460s &  .658 &  .641 &  .649 &  6.342s &  .563 &  .577 &  .570 &  5.590s &  .713 &  .753 &  .732 \\ \hline
ESP + 1st & 7.118s &  .814 &  .797 &  \textbf{.805} &  8.301s &  .654 &  .675 &  \textbf{.664} &  7.837s &  .867 &  .916 & .891 \\ \hline
\end{tabular}
\caption{Result for Class D, X and A. Crowd-powered entity extraction outperforms the CRF baseline in terms of F1-score on both Class D and X queries.
Although the CRF baseline is well-developed on Class A,
it is not effective on complex queries.}
\label{tab:classDX}
\end{table*}

Based on the study above,
for Class D and X queries,
we use the Dialog ESP Game of 10 players with ``ESP+1st'' and ``1st Only'' settings 
to measure the best F1-score and speed.
The time constraint is set to 20 seconds.
The experiments are run on MTurk and all settings are identical as the previous section.
76 distinct workers participate in Class D experiments, and 68 distinct workers participate in Class X experiments.

Experimental results are shown in Table~\ref{tab:classDX}.
An automated CRF model is implemented as a baseline.\footnote{Implemented with CRF++: http://taku910.github.io/crfpp/}
The CRF model is trained on the Class-A training set mentioned above by using neighbor words (window size $=$ 2) and POS tag features.
The CRF model is decoded and timed on a laptop with Intel i5-4200U CPU (@1.60GHz) and 8GB RAM.
As a result, the proposed crowd-powered approach largely outperforms the CRF baseline in terms of F1-score on both Class D and X queries .
Although the CRF approach is well-developed on Class A data,
it is not effective on the remaining data.

Surprisingly, we find similar average response times in each query category.
Note that the text length is different for each category:
the average token number of Class-A queries is 11.47, 
of Class-D queries (including the query history) is 48.64,
and of Class-X queries is 67.72.
Studies showed that eyes' warm-up time~\cite{FirstFix} and word frequency influence speed of text comprehension~\cite{ReadingSpeedWordFrequency,ReadingSpeedWordFrequencyClassic}.
These factors might reduce the effect of text length to the reading speed of crowd works.

\begin{table}[t]
\centering
\small
\begin{tabular}{@{}lrrr@{}}
\toprule
\multicolumn{1}{c}{Error Type} & \multicolumn{1}{c}{Class D} & \multicolumn{1}{c}{Class X} & \multicolumn{1}{c}{Class A} \\ \midrule
\texttt{fromloc.city\_name} & 39.53\% & 16.67\% & 40.00\% \\
False Negative & 18.60\% & 26.67\% & 0.00\% \\
Incorrect City & 16.28\% & 18.33\% & 8.00\% \\
Correct City \& Soft Match & 16.28\% & 5.00\% & 12.00\% \\
False Positive & 9.30\% & 33.33\% & 40.00\% \\ \bottomrule
\end{tabular}
\caption{Error Analysis for Class D, X and A.}
\label{tab:error}
\end{table}

We also conduct an error analysis on the result of ``ESP+1st'' setting, which achieves our best F1-score.
The distribution of error types are shown in Table~\ref{tab:error}.
The ``\texttt{fromloc.city\_name}'' type indicates that the crowd extracts the departure city, rather than destination city;
In ``Incorrect City'' type, the crowd extracts an incorrect city from the query history (but not the departure city);
``Correct City \& Soft Match'' type means the extracted city name is semantically correct but does not match the gold-standard city name
(e.g., ``Washington'' and ``Washington DC'').
From the error analysis, we conclude two directions to improve performance: 
1) treat the cases of absent slot more carefully, and 
2) use domain knowledge if available.
First, 28\% of errors in Class D and 50\% in Class X occur when either the gold-standard label or the predicted label does not exist.
It suggests that a more reliable step to recognize the existence of the targeted entity might be required.
Second, 16.28\% of Class-D queries and 5\% of Class-X queries are of the ``Soft Match'' cases.
By introducing domain knowledge like a list of city names,
a post-processor that finds the most similar city name of the predicted label can fix this type of error.

\section{Experiments 2:\\User Experiment via a Real-world Instant Messaging Interface}

To examine the feasibility of real-time crowd-powered entity extraction in an actual system, we conduct lab-based user experiments via Google Hangouts' instant messaging interface.
Our proposed method has a task completion time of 5-8 seconds, per Experiment 1.
In this section, we demonstrate our approach is robust and fast enough to support a real-world instant messaging application,
where the average time gap between conversational turns is 24 seconds~\cite{isaacs2002character}.

\subsection{System Implementation}

\begin{figure*}[htbp]
    \centering
    \includegraphics[width=\textwidth]{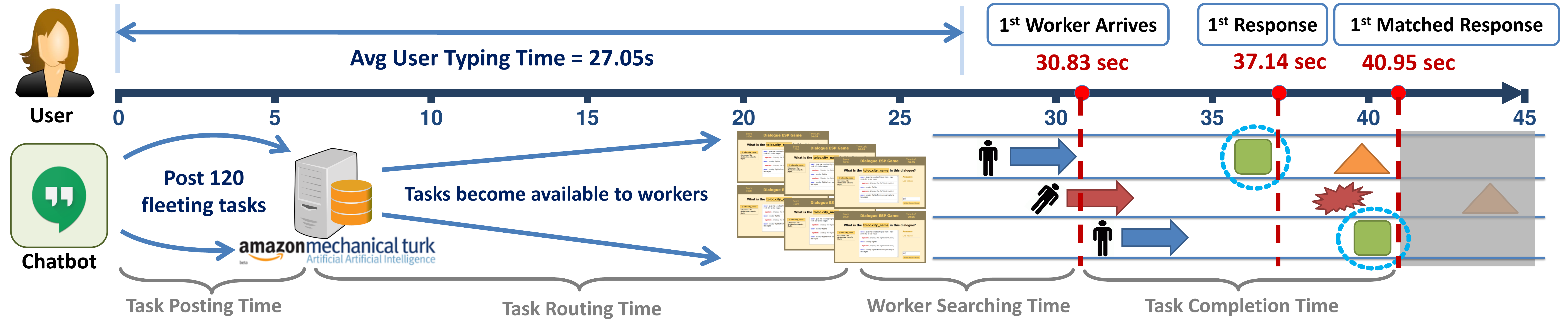}
    \caption{Timeline of the Real-time Crowd-powered Entity Extraction System. On average, the first worker takes 30.83 seconds to reach, 
    the first answer is received at 37.14 seconds,
    and the first matched answer occurs at 40.95 seconds.
    A user on average spends 27.05 seconds to type a chat line, i.e., the perceived response time to users falls within 10-14 seconds.
    }
    \label{fig:google-system}
\end{figure*}

We implemented a Google Hangouts chatbot by using the Hangupsbot\footnote{https://github.com/hangoutsbot/hangoutsbot} framework.
Users are able to send text chats to our chatbot via Google Hangouts.
The chatbot recruits crowd workers on MTurk in real-time to perform the Dialog ESP Game task upon receiving the chat.
Figure~\ref{fig:google-system} shows the overview of our system.
We record all answers submitted by recruited workers and log the timestamps of following activities:
1) users' and workers' keyboard typing, 
2) workers' task arrival, and 
3) the workers' answer submissions.

To recruit crowd workers, we introduce \textit{fleeting task}, a recruiting practice inspired by \textit{quikturkit}~\cite{VizWiz2010}.
This approach achieves low latency by posting hundreds of short lifetime tasks, which increases task visibility.
Its short lifetime (e.g., 60 seconds) encourages workers to complete tasks quickly.
A core benefit of the \textit{fleeting task} approach is its ease in implementation:
the method bypasses the common practices of pre-recruiting workers and maintaining a waiting pool~\cite{VizWiz2010,Chorus2013,CrowdIn2Sec2011}.
In a system deployed at scale, a retainer or push model is likely to work as well.


\subsection{User Experiment Setup}

We conduct lab-based user experiments to evaluate the proposed technology on extracting ``food'' entities.
Ten Google Hangouts users enter our lab with their own laptops.
We first ask them to arbitrarily create a list 9 foods, 3 drinks, and 3 countries based on their own preferences.
Then we explain the purpose of the experiments, and introduce five scenarios of using instant messaging:

\begin{enumerate}
    \item \textbf{Eat}:
        You discuss with your friend about what to eat later.
    \item \textbf{Drink}:
        You discuss with an employee a coffee place, bar, or restaurant to order something to drink.
    \item \textbf{Cook}:
        You plan to cook later. You discuss the details with your friend who knows how to cook.
    \item \textbf{Chat}:
        You are chatting with your friend.
    \item \textbf{No Food}:
        You are chatting with your friend.
        You do not mention food.
        Instead, you mention a country name.
\end{enumerate}

We also list three types of conversational acts which could emerge in each scenario:

\begin{enumerate}
    \item \textbf{Question}:
        Ask a question.
    \item \textbf{Answer}:
        Answer a question that could be asked under the current scenario.
    \item \textbf{Mentioning}:
        Naturally converse without asking or answering any specific questions. 
\end{enumerate}

Using their laptops, users send one text chat for each combination of [scenario, conversational act] to our chatbot, i.e., 15 chats in total.
In the Eat, Cook, and Chat scenarios, users must mention one of the foods they listed earlier;
in the Drink scenario, they must mention one of the drinks they listed.
In the No Food scenario, users must mention one of the countries they listed, and no food names can be mentioned.
In total, we collect 150 chat inputs from 10 user experiments.
Correspondingly, instructions on the workers' interface (Figure~\ref{fig:ui}) is modified as ``What is the \texttt{food\_name} in this dialog?'',
and the explanation of \texttt{food\_name} is modified as ``\textit{Food name. The full name of the food. Including any drinks or beverages.}''
In the experiments, our chatbot post 120 HITs with a lifetime of 60 seconds to MTurk upon receiving a text chat.
The price of each HIT is \$0.1. 
We use the interface shown in Figure~\ref{fig:ui} with a time constraint of 20 seconds. 

\subsection{Experimental Results}

\begin{table}[t]
\centering
\small
\begin{tabular}{@{}ccc@{}}
\toprule
 & \textbf{Acc (\%)} & \textbf{\begin{tabular}[c]{@{}c@{}}Response Time (s)\\ Mean (Stdev)\end{tabular}} \\ \midrule
\textbf{1st Only} & 77.33\% & 37.14 (14.70) \\ \midrule
\textbf{ESP Only} & 81.33\% & 40.95 (13.56) \\ \midrule
\textbf{ESP + 1st} & 84.00\% & 40.95 (13.56) \\ \midrule
\multicolumn{2}{c}{\textbf{1st Worker Reached Time (s)}} & 30.83 (16.86) \\ \midrule
\multicolumn{2}{c}{\textbf{User Type Time (s)}} & 27.05 (25.28) \\ \bottomrule
\end{tabular}
\caption{Result of User Experiment. A trade-off between time and output quality can be observed.}
\label{tab:google-result}
\end{table}

Results are shown in Table~\ref{tab:google-result}.
The ``ESP+1st'' setting achieves the best accuracy of 84\% with an average response time of 40.95 seconds.
The ``1st Only'' setting has the shortest average response time of 37.14 seconds with an accuracy of 77.33\%.\footnote{We only consider the answers submitted within 60 seconds.}
A trade-off between time and output quality can be observed.
This trade-off is similar to the results of Experiment 1 (shown in Figure~\ref{fig:result}(c)).
On average, 14.45 MTurk workers participated in each trial and submitted 33.81 answers.

\begin{table}[t]
\centering
\small
\begin{tabular}{@{}cccccc@{}}
\toprule
\multicolumn{2}{c}{\multirow{2}{*}{}} & \multicolumn{2}{c}{1st} & \multicolumn{2}{c}{ESP + 1st} \\ \cmidrule(l){3-6} 
\multicolumn{2}{c}{} & \begin{tabular}[c]{@{}c@{}}Avg. \\ Time(s)\end{tabular} & \begin{tabular}[c]{@{}c@{}}Acc.\\ (\%)\end{tabular} & \begin{tabular}[c]{@{}c@{}}Avg.\\ Time(s)\end{tabular} & \begin{tabular}[c]{@{}c@{}}Acc.\\ (\%)\end{tabular} \\ \midrule
\multirow{3}{*}{\begin{tabular}[c]{@{}c@{}}Entity\\ Type\end{tabular}} & Food\footnotemark & 36.64 & 70.00\% & 40.19 & \textbf{78.89\%} \\ \cmidrule(l){2-6} 
 & Drink & 37.43 & 80.00\% & 41.37 & \textbf{83.33\%} \\ \cmidrule(l){2-6} 
 & None & 38.33 & 96.67\% & 42.83 & \textbf{100.00\%} \\ \midrule
\multirow{3}{*}{\begin{tabular}[c]{@{}c@{}}Conv.\\ Act\end{tabular}} & Question & 34.26 & 82.00\% & 37.94 & \textbf{90.00\%} \\ \cmidrule(l){2-6} 
 & Answer & 39.90 & 68.00\% & 43.88 & \textbf{78.00\%} \\ \cmidrule(l){2-6} 
 & Mention & 37.26 & 82.00\% & 41.04 & \textbf{84.00\%} \\ \midrule
\multicolumn{2}{c}{\textbf{Avg.}} & \textbf{37.14} & \textbf{77.33\%} & \textbf{40.95} & \textbf{84.00\%} \\ \bottomrule
\end{tabular}
\caption[Caption for LOF]{Results of user experiment for each scenario and conversational act.}
\label{tab:scenario}
\end{table}

\footnotetext{Including the results from Food, Cook, and Chat scenarios.}

\paragraph{Robustness in Out-of-Vocabulary Entities \& Language Variability}

The results over each entity type are shown in Table~\ref{tab:scenario}.
Without using any training data or pre-defined knowledge-base, our crowdsourcing approach achieves an accuracy of 78.89\% in extracting food entities and 83.33\% in extracting drink entities.
Despite the significant variety of the input entities\footnote{
The food entities arbitrarily created by our users are quite diverse:
From a generic category (e.g., Thai food) to a specific entry (e.g., Magic Hat \#9), and
from a simple food (e.g., cherry) to a complex food (e.g., sausage muffin with egg).
The list covers the food of 
many other countries (e.g., Okonomiyaki, Bibimbap, Samosa.)
}, our approach extracts most entities correctly.
Furthermore, our method is effective in identifying the absence of entities;
Table~\ref{tab:scenario} also shows the robustness of the proposed method under various linguistic conditions.
The ``ESP+1st'' setting achieves accuracies of 90.00\% in extracting entities from questions,
78.00\% in extracting from answers,
and 84.00\% in extracting from regular conversations.
Qualitatively, our approach can handle complex input, such as strange restaurant names and beverage names, which are essentially confusing for automated approaches.
For example, ``Have you ever tried bibimbap at \textit{Green pepper}?'' and 
``I usually have \textit{Magic Hat \#9}'', where \textit{Green pepper} and \textit{Magic Hat \#9} are names of a restaurant and beverage, respectively.

\paragraph{Error Analysis}

Table~\ref{tab:error-google} shows the errors in the user experiments (``ESP+1st'' setting).
45.83\% of errors are caused by absence of answers, mainly due to the task routing latency of the MTurk platform.
We discuss this in more detail below.
37.50\% of errors are due to various system problems such as the string encoding issues.
More interestingly, 12.50\% of incorrect answers are sub-spans of the correct answers.
For instance, the crowd extracts ``rice'' for ``stew pork over rice'',
and ``tea'' for ``bubble tea''.
This type of error is similar to the ``Soft Match'' error in Experiment 1.
Finally, 4.17\% of errors are caused by user typos (e.g., \textit{latter} for \textit{latte}), which the crowd tends to exclude in their answers.

\begin{table}[htbp]
\small
\centering
\begin{tabular}{@{}ll@{}}
\toprule
\multicolumn{1}{c}{\textbf{Error Type}} & \multicolumn{1}{c}{\textbf{\%}} \\ \midrule
No Answers Received & 45.83\% \\
System Problem & 37.50\% \\
Substring of a Multi-token Entity & 12.50\% \\
Typo & 4.17\% \\ \bottomrule
\end{tabular}
\caption{Error Analysis for User Experiment.}
\label{tab:error-google}
\end{table}

\paragraph{Response Speed}
Table~\ref{tab:google-result} shows the average response time in the user experiment.
On average, the first worker takes 30.83 seconds to reach to our Dialog ESP Game,
the first answer is received at 37.14 seconds,
and the first matched answer occurs at 40.95 seconds. 
For comparison, we illustrate the timeline of our system in Figure~\ref{fig:google-system}.
In the user experiments, a user on average spends 27.05 seconds to type a chat line.
If we align the user typing time along with the system timeline, 
the theoretical perceived response time to users falls within 10-14 seconds,
while the average response time in instant messaging is 24 seconds~\cite{isaacs2002character}.
\cite{baron2010discourse} reports that 24.5\% of instant messages get responses within 11-30 seconds,
and 8.2\% of messages have even longer response times.
The proposed technology proves to be fast enough to support instant messaging applications.
The main bottleneck of the end-to-end response speed is the \textit{task routing time} in Figure~\ref{fig:google-system},
which approximately ranges from 5-40 seconds and changes over time.
The task routing time also causes the major errors in Table~\ref{tab:error-google}.
The task lifetime begins when a task reaches the MTurk server instead of when it becomes visible to workers.
When the task routing time is longer than a task's lifetime, 
the task could expire before it is selected by workers.
Because MTurk requesters can not effectively reduce the task routing time,
pre-recruiting and queuing workers seems inevitable for applications which require a response time sharply shorter than 30 seconds.




\section{Discussion}



Incorporating domain-specific knowledge is a major obstacle in generalization of crowdsourcing technologies~\cite{huang2015guardian}.
We think that automation helps resolve this challenge.
One most common errors in our system are the \textit{soft match}, where the
crowd extracts a sub-string of the target entity instead of the complete string.
Domain knowledge can help to fix this type of errors.
However, unlike automated technology, we do not have a generic method to update human workers with new knowledge.
Thus, our next step is to incorporate automated components.
It is easy to replace some workers with automated annotators in our multi-player ESP Game.
Despite fragility in extracting unseen entities, automated approaches are robust in identifying known entities and can be easily updated if new data is collected.
We will develop a hybrid approach,
which we believe will be robust in unexpected input and easily incorporate new knowledge.



\section{Conclusion and Future Work}


We have explored using real-time crowdsourcing to extract entities for dialog systems.
By using an ESP Game setting, our approach is absolute 36.5\% and 27.1\% better than the CRF baseline 
in terms of F1-score for Class D and X queries in the ATIS dataset, respectively.
The timing cost is about 8 seconds, which is slower than machines but still reasonable given the large gains in accuracy.
The proposed method also has been evaluated via Google Hangouts' text chat with 10 users.
The results demonstrate the robustness and feasibility of our approach in real-world systems.
In the future, we will generalize our approach
by adding automated components, and also explore the possibility of using audio input.

\section{Appendix: List of Food and Drinks Used in the Experiment 2}

The followings are the lists of 9 food and 3 drinks created by 10 participants in our user experiment.

\subsection{Food}

\begin{enumerate}
    \item spaghetti, burger, vindaloo lamb,
makhani chicken, kimchee, wheat bread
pizza, cornish pasty, mushroom soup
\vspace{-.4pc} \item burger, french fries, scallion cake,
okonomiyaki, oyakodon, gyudon,
fried rice, wings, salad
\vspace{-.4pc} \item Stinky Tofu, Acai Berry Bowl, Tuna Onigiri,
Rice Burger, Seared Salmon, Milkfish Soup,
Mapo Tofu, Beef Pho, Scallion Pancake
\vspace{-.4pc} \item pizza, fried rice, waffle,
alcohol drink, chocolate pie, cookie,
dimsum, burger, milk shake
\vspace{-.4pc} \item Pho, BBQ, Thai food, 
beef noodles, steak, Tomato soup,
Spicy hot pot, Soup dumplings, Ramen
\vspace{-.4pc} \item chocolate, donut, cheesecake,
pad thai, seafood pancake, fish fillets in hot chili,
hot pot, bibimbap, japchae
\vspace{-.4pc} \item chocolate, pancakes, strawberries,
fried fish, fried chicken, sausages,
gulaab jamun, paneer tika, samosa
\vspace{-.4pc} \item Dumplings, noodle, stew pork over rice,
Sandwich, pasta, hot pot,
Potato slices with green peppers, Chinese BBQ, pancakes
\vspace{-.4pc} \item stinky tofu, stew pork over rice, yakitori,
baked cinnamon apple, apple pie, stew pork with potato and apple,
teppanyaki, okonomiyaki, crab hotpot
\vspace{-.4pc} \item hot pot, cherry, Chinese cabbage,
Pumpkin risotto, Tomato risotto, Boeuf Bourguignon,
stinky tofu, sausage muffin with egg (McDonald), eggplant with basil
\end{enumerate}

\subsection{Drink}

\begin{enumerate}
    \item tea, coke, latte
\vspace{-.4pc} \item green tea latte, bubble tea, root beer
\vspace{-.4pc} \item medium latte with non-fat milk, green Tea Latte,\\Soymilk
\vspace{-.4pc} \item water, pepsi, tea
\vspace{-.4pc} \item Latte with nonfat milk, Magic hat \#9, Old fashion
\vspace{-.4pc} \item vanilla latte, strawberry smoothie, iced tea
\vspace{-.4pc} \item coffee, milk shake, beer
\vspace{-.4pc} \item Mocha coffee, beers, orange juice
\vspace{-.4pc} \item caramel frappuccino, caramel macchiato, coffee with coconut milk
\vspace{-.4pc} \item ice tea, macha, apple juice
\end{enumerate}

\bigskip

\bibliographystyle{aaai}
\bibliography{homp_2016}

\end{document}